\documentclass[11pt]{article}
\usepackage{utopia}
\usepackage[active]{srcltx}
\usepackage{multicol}
\usepackage{graphicx}
\usepackage{amssymb}
\usepackage{amsthm}
\title{\bf Two-Level Laser Dynamics with a Noiseless Vacuum Reservoir}
\author{Fesseha Kassahun\footnote{Email address: fesseha.kassahun@aau.edu.et} \\
                \footnotesize{Department of Physics, Addis Ababa University, P. O. Box 33761, Addis Ababa, Ethiopia}}
                \date{\footnotesize{(Submitted on 21 Sep 2012)}}

\makeatother

\begin{document}
\maketitle
\begin{abstract}
We analyze the quantum properties of the light generated by a two-level laser in which the two-level atoms available in a closed cavity are pumped to the upper level by means of electron bombardment. We consider the case in which the two-level laser is coupled to a vacuum reservoir via a single-port mirror and seek to carry out our analysis by putting the noise operators associated with the vacuum reservoir in normal order. It is found that the two-level laser generates coherent light when operating well above threshold and chaotic light when operating at threshold.
Moreover, we have established that a large part of the total mean photon number is confined in a relatively small frequency interval. This is a revised version of a paper published in Opt. Commun. (284, 1357, 2011).
\end{abstract}
\hspace*{9.5mm}Keywords: Photon statistics, Quadrature variance, Power spectrum
\vspace*{3mm}

\section{Introduction}
A two-level laser is a source of coherent or chaotic light emitted by two-level atoms inside a cavity coupled to a vacuum reservoir via a single-port mirror. In one model of such a laser, two-level atoms initially in the upper level are injected at a constant rate into a cavity and removed after they have decayed due to spontaneous emission [1,2]. In another model the two-level atoms available in a cavity are pumped to the upper level by some convenient means such as electron bombardment [2,3].

There has been a considerable interest to study the quantum properties of the light generated by a two-level laser [4-10]. It is found that the light generated by this laser operating well above threshold is coherent and the light generated by the same laser operating below threshold is chaotic. In the quantum theory of a laser, one usually takes into consideration the interaction of the atoms inside the cavity with the vacuum reservoir outside the cavity. There may be some justification for the possibility of such interaction for a laser with an open cavity into which and from which atoms are injected and removed. However, there cannot be any valid justification for leaving open the laser cavity in which the available atoms are pumped to the upper level by means of electron bombardment. Therefore, the aforementioned interaction is not feasible for a laser in which the atoms available in a closed cavity are pumped to the upper level by electron bombardment.

We seek here to analyze the quantum properties of the light emitted by two-level atoms available in a closed cavity and pumped to the upper level at a constant rate. Thus taking into account the interaction of the two-level atoms with a resonant cavity mode and the damping of the cavity mode by a vacuum reservoir,
we obtain the photon statistics, the quadrature variance, and the power spectrum for the light emitted by the atoms. We carry out this analysis by putting the noise operators associated with the vacuum reservoir
in normal order and without considering the interaction of the two-level atoms with the vacuum reservoir outside the cavity.

\section{Operator dynamics}
We consider here the case in which $N$ two-level atoms are available in a closed cavity.
Then the interaction of the cavity mode with one of the atoms can be described
at resonance by the Hamiltonian
\begin{equation}\label{1}\hat{H}=ig(\hat{\sigma}_{a}^{\dag k}\hat{b}-\hat{b}^{\dag}
\hat{\sigma}_{a}^{k}), \end{equation}
where
\begin{equation}\label{2}
\hat{\sigma}^{k}_{a}=|b\rangle_{k}~_{k}\langle a|
\end{equation}
is a lowering atomic operator, $\hat{b}$ is the
annihilation operator for the cavity mode, and $g$ is the coupling constant between the atom and the cavity mode.
We assume that the laser cavity is coupled to a vacuum reservoir via a single-port mirror. In addition, we carry out our calculation by putting the noise operators associated with the vacuum reservoir in normal order. Thus the noise operators will not have any effect on the dynamics of the cavity mode operators. We can then drop the noise operator and write the quantum Langevin equation for the operator $\hat{b}$ as
\begin{equation}\label{3}
{d\hat{b}\over dt}=-{\kappa\over 2}\hat{b}-i[\hat{b},\hat{H}],
\end{equation}
where $\kappa$ is the cavity damping constant. Therefore, with the aid of (\ref{1}), we readily find
\begin{equation}\label{4}
{d\hat{b}\over dt}=-{\kappa\over 2}\hat{b}-g\hat{\sigma}_{a}^{k}.
\end{equation}

Furthermore, employing the relation
\begin{equation}\label{5}
{d\over dt}\langle\hat{A}\rangle=-i\langle[\hat{A},\hat{H}]\rangle
\end{equation}
along with Eq. (\ref{1}), one readily obtains
\begin{equation}\label{6}
{d\over dt}\langle\hat{\sigma}_{a}^{k}\rangle=g\langle(\hat{\eta}_{b}^{k}-\hat{\eta}_{a}^{k})\hat{b}\rangle, \end{equation}
\begin{equation}\label{7}
{d\over dt}\langle\hat{\eta}_{a}^{k}\rangle=g\langle\hat{\sigma}_{a}^{\dag k}\hat{b}\rangle +g\langle\hat{b}^{\dag}\hat{\sigma}_{a}^{k}\rangle,
\end{equation}
\begin{equation}\label{8}
{d\over dt}\langle\hat{\eta}_{b}^{k}\rangle=-g\langle\sigma_{a}^{\dagger k}\hat{b}\rangle-g\langle\hat{b}^{\dagger}\sigma_{a}^{k}\rangle,
\end{equation}
where
\begin{equation}\label{9}
\hat{\eta}_{a}^{k}=|a\rangle_{k}~_{k}\langle a|,
\end{equation}
\begin{equation}\label{10}
\hat{\eta}_{b}^{k}=|b\rangle_{k}~_{k}\langle b|.
\end{equation}

We see that Eqs. (\ref{6}), (\ref{7}), and (\ref{8}) are nonlinear differential equations and hence it is not possible to obtain exact time-dependent solutions of these equations. Thus applying  the large-time approximation scheme [11], we obtain from Eq. (\ref{4}) the approximately valid relation
\begin{equation}\label{11}
\hat{b}(t)=-{2g\over\kappa}\hat{\sigma}^{k}_{a}(t).
\end{equation}
Now substitution of (\ref{11}) (with the time argument suppressed) into the aforementioned equations yields
\begin{equation}\label{12}
{d\over dt}\langle\hat{\sigma}_{a}^{k}\rangle=-{1\over 2}\gamma_{c}\langle\hat{\sigma}_{a}^{k}\rangle,
\end{equation}
\begin{equation}\label{13}
{d\over dt}\langle\hat{\eta}_{a}^{k}\rangle=-\gamma_{c}\langle\hat{\eta}_{a}^{k}\rangle,
\end{equation}
\begin{equation}\label{14}
{d\over dt}\langle\hat{\eta}_{b}^{k}\rangle=\gamma_{c}\langle\hat{\eta}_{a}^{k}\rangle,
\end{equation}
where
\begin{equation}\label{15}\gamma_{c}=4g^{2}/\kappa.
\end{equation}
We wish to call the parameter defined by Eq. (\ref{15}) the stimulated emission decay constant.
Based on the definition of this decay constant, we infer that an atom in the upper level and inside a closed cavity  emits a photon due to its interaction with the cavity light. We certainly identify this process to be stimulated photon emission.

In order to include the contribution of all the atoms to the dynamics of the two-level laser, we sum Eqs. (\ref{12}), (\ref{13}), and (\ref{14}) over the $N$ two-level atoms, so that
\begin{equation}\label{16}
{d\over dt}\langle\hat{m}_{a}\rangle=-{1\over 2}\gamma_{c}\langle\hat{m}_{a}\rangle,
\end{equation}
\begin{equation}\label{17}
{d\over dt}\langle\hat{N}_{a}\rangle=-\gamma_{c}\langle\hat{N}_{a}\rangle,
\end{equation}
\begin{equation}\label{18}
{d\over dt}\langle\hat{N}_{b}\rangle=\gamma_{c}\langle\hat{N}_{a}\rangle,
\end{equation}
in which
\begin{equation}\label{19}
\hat{m}_{a}=\sum_{k=1}^{N}\hat{\sigma}_{a}^{k},
\end{equation}
\begin{equation}\label{20}
\hat{N}_{a}=\sum_{k=1}^{N}\hat{\eta}_{a}^{k},
\end{equation}
\begin{equation}\label{21}
\hat{N}_{b}=\sum_{k=1}^{N}\hat{\eta}_{b}^{k},
\end{equation}
with the operators $\hat{N}_{a}$ and $\hat{N}_{b}$ representing the number of atoms in the upper and lower levels.

Furthermore, with the aid of the identity
\begin{equation}\label{22}
\hat{\eta}_{a}^{k}+\hat{\eta}_{b}^{k}=\hat{I},
\end{equation}
we see that
\begin{equation}\label{23}
\langle\hat{N}_{a}\rangle+\langle\hat{N}_{b}\rangle=N.
\end{equation}
In addition, using the definition given by (\ref{2}) and setting for any $k$
\begin{equation}\label{24}
\hat{\sigma}_{a}^{\kappa}=|b\rangle\langle a|,
\end{equation}
we have
\begin{equation}\label{25}
\hat{m}_{a}=N|b\rangle\langle a|.
\end{equation}
We therefore observe that
\begin{equation}\label{26}
\hat{m}_{a}^{\dagger}\hat{m}_{a}=N\hat{N}_{a},
\end{equation}
in which
\begin{equation}\label{27}
\hat{N}_{a}=N|a\rangle\langle a|.
\end{equation}
Following the same procedure, one can also establish that
\begin{equation}\label{28}
\hat{m}_{a}\hat{m}_{a}^{\dagger}=N\hat{N}_{b},
\end{equation}
with
\begin{equation}\label{29}
\hat{N}_{b}=N|b\rangle\langle b|.
\end{equation}

In the presence of $N$ two-level atoms, we rewrite Eq. (\ref{4}) as
\begin{equation}\label{30}
{d\hat{b}\over dt}=-{\kappa\over 2}\hat{b}+\lambda\hat{m}_{a}.
\end{equation}
We now proceed to determine the value of the constant $\lambda$. Thus applying the steady-state solution of Eq. (\ref{4}), we easily find
\begin{equation}\label{31}
[\hat{b},\hat{b}^{\dagger}]_{k}={\gamma_{c}\over\kappa}(\hat{\eta}_{b}^{k}-\hat{\eta}_{a}^{k})
\end{equation}
and on summing over all atoms, we have
\begin{equation}\label{32}
[\hat{b},\hat{b}^{\dagger}]={\gamma_{c}\over\kappa}(\hat{N}_{b}-\hat{N}_{a}),
\end{equation}
where
\begin{equation}\label{33}
[\hat{b},\hat{b}^{\dagger}]=\sum_{k=1}^{N}[\hat{b},\hat{b}^{\dagger}]_{k}
\end{equation}
stands for the commutator of $\hat{b}$ and $\hat{b}^{\dagger}$ when the cavity mode is interacting with all the $N$ two-level atoms.
On the other hand, using the steady-state solution of Eq. (\ref{30}), one can easily establish that
\begin{equation}\label{34}
[\hat{b},\hat{b}^{\dagger}]=N\bigg({2\lambda\over\kappa}\bigg)^{2}(\hat{N}_{b}-\hat{N}_{a}).
\end{equation}
Hence from Eqs. (\ref{32}) and (\ref{34}), we get
\begin{equation}\label{35}
\lambda={\pm}{g\over\sqrt{N}}
\end{equation}
and on account of this result, Eq. (\ref{30}) can be written as
\begin{equation}\label{36}
{d\hat{b}\over dt}=-{\kappa\over 2}\hat{b}+{g\over\sqrt{N}}\hat{m}_{a}.
\end{equation}

The two-level atoms available in the cavity are pumped from the lower to the upper level by means of electron bombardment. The pumping process must certainly affect the time evolution of the atomic operators. Hence we take into account the effect of the pumping process on the time evolution of the operators $\langle\hat{N}_{a}\rangle$ and $\langle\hat{N}_{b}\rangle$ by rewriting Eqs. (\ref{17}) and (\ref{18}) as
\begin{equation}\label{37}
{d\over dt}\langle\hat{N}_{a}\rangle=-\gamma_{c}\langle\hat{N}_{a}\rangle+r_{a}\langle\hat{N}_{b}\rangle
\end{equation}
and
\begin{equation}\label{38}
{d\over dt}\langle\hat{N}_{b}\rangle=-r_{a}\langle\hat{N}_{b}\rangle+\gamma_{c}\langle\hat{N}_{a}\rangle,
\end{equation}
where $r_{a}$ is the rate at which a single atom is pumped to the upper level.
Now taking into account (\ref{23}), one can put Eq. (\ref{37}) in the form
\begin{equation}\label{39}
{d\over dt}\langle\hat{N}_{a}\rangle=-(\gamma_{c}+r_{a})\langle\hat{N}_{a}\rangle+r_{a}N.
\end{equation}
We immediately see that the steady-state solution of this equation is
\begin{equation}\label{40}
\langle\hat{N}_{a}\rangle={r_{a}N\over{\gamma_{c}+r_{a}}}
\end{equation}
and the steady-state solution of Eq. (\ref{38}) turns out to be
\begin{equation}\label{41}
\langle\hat{N}_{b}\rangle={\gamma_{c}\over{r_{a}}}\langle\hat{N}_{a}\rangle.
\end{equation}

Finally, we seek to determine the effect of the pumping process on the dynamics of the atomic operator $\hat{m}_{a}$. To this end, we rewrite Eq. (\ref{16}) as
\begin{equation}\label{42}
{d\over dt}\hat{m}_{a}=-{1\over 2}\eta\hat{m}_{a}+\hat{F}_{a}(t),
\end{equation}
where $\hat{F}_{a}(t)$ is a noise operator with a vanishing mean and $\eta$ is a parameter whose value remains to be fixed. Employing the relation
\begin{equation}\label{43}
{d\over dt}\bigg\langle\hat{m}_{a}^{\dagger}\hat{m}_{a}\bigg\rangle=\bigg\langle{d\hat{m}_{a}^{\dagger}\over dt}\hat{m}_{a}\bigg\rangle+\bigg\langle\hat{m}_{a}^{\dagger}{d\hat{m}_{a}\over dt}\bigg\rangle
\end{equation}
along with Eqs. (\ref{42}) and (\ref{26}), we get
\begin{equation}\label{44}
{d\over dt}\langle\hat{N}_{a}(t)\rangle=-\eta\langle\hat{N}_{a}(t)\rangle+{1\over N}\langle\hat{F}_{a}^{\dagger}(t)\hat{m}_{a}(t)\rangle
+{1\over N}\langle\hat{m}_{a}^{\dagger}(t)\hat{F}_{a}(t)\rangle.
\end{equation}
Now comparison of Eqs. (\ref{39}) and (\ref{44}) shows that
\begin{equation}\label{45}
\eta=\gamma_{c}+r_{a}
\end{equation}
and
\begin{equation}\label{46}
\langle\hat{F}_{a}^{\dagger}(t)\hat{m}_{a}(t)\rangle+\langle\hat{m}_{a}^{\dagger}(t)\hat{F}_{a}(t)\rangle=r_{a}N^{2}.
\end{equation}
This result implies that
\begin{equation}\label{47}
\langle\hat{F}_{a}^{\dagger}(t)\hat{F}_{a}(t')\rangle=r_{a}N^{2}\delta (t-t').
\end{equation}
Following a similar procedure, one can also easily establish that
\begin{equation}\label{48}
\langle\hat{F}_{a}(t)\hat{F}_{a}^{\dagger}(t')\rangle=\gamma_{c}N^{2}\delta (t-t').
\end{equation}

\section{Photon statistics}
It is certainly interesting to consider different regimes of the laser operation. We then wish to call the regime of laser operation with more atoms in the upper level than in the lower level above threshold, and the regime of laser operation with equal number of atoms in the upper and lower levels threshold. Thus inspection of Eq. (\ref{41}) shows that for the laser operating at threshold $\gamma_{c}=r_{a}$ and for the laser operating above threshold $\gamma_{c}<r_{a}$.
Applying the solution of Eq. (\ref{42}) and considering the atoms to be initially in the lower level, we easily find
\begin{equation}\label{49}
\langle\hat{m}_{a}(t)\rangle=0.
\end{equation}
Moreover, the expectation value of the solution of Eq. (\ref{36}) is expressible as
\begin{equation}\label{50}
\langle\hat{b}(t)\rangle=\langle\hat{b}(0)\rangle e^{-\kappa t/2}
+{g\over\sqrt{N}}e^{-\kappa t/2}
\int_{0}^{t}e^{\kappa t'/2}\langle\hat{m}_{a}(t')\rangle dt'.
\end{equation}
On account of (\ref{49}) and the assumption that the cavity light is initially in a vacuum state,  Eq. (\ref{50}) reduces to
\begin{equation}\label{51}
\langle\hat{b}(t)\rangle=0.
\end{equation}
We note from Eqs. (\ref{36}) and (\ref{51}) that $\hat{b}$ is a Gaussian variable with zero mean.

Using the steady-state solution of Eq. (\ref{36}),
\begin{equation}\label{52}
\hat{b}={2g\over\kappa\sqrt{N}}\hat{m},
\end{equation}
we get
\begin{equation}\label{53}
\langle\hat{b}^{\dagger}\hat{b}\rangle={\gamma_{c}\over\kappa}\langle\hat{N}_{a}\rangle,
\end{equation}
so that in view of (\ref{40}), the mean photon number has at steady state the form\index{mean photon number!for a two-level laser}
\begin{equation}\label{54}
\overline{n}={\gamma_{c}\over\kappa}\bigg({r_{a}N\over{\gamma_{c}+r_{a}}}\bigg).
\end{equation}
We note that for the two-level laser operating well above threshold $(\gamma_{c}\ll r_{a})$,  Eq. (\ref{54}) reduces to
\begin{equation}\label{55}
\overline{n}={\gamma_{c}\over\kappa}N.
\end{equation}
And for the same laser operating at threshold $(\gamma_{c}=r_{a})$, we have
\begin{equation}\label{56}
\overline{n}={\gamma_{c}\over 2\kappa}N.
\end{equation}

Furthermore, the variance of the photon number for the cavity light can be written as
\begin{equation}\label{57}
(\Delta n)^{2}=\langle\hat{b}^{\dagger}\hat{b}\hat{b}^{\dagger}\hat{b}\rangle
-\langle\hat{b}^{\dagger}\hat{b}\rangle^{2} \end{equation}
and recalling that $\hat{b}$ is a Gaussian variable with zero mean , we readily get
\begin{equation}\label{58}
(\Delta n)^{2}=\langle\hat{b}^{\dagger}\hat{b}\rangle\langle\hat{b}\hat{b}^{\dagger}\rangle.
\end{equation}
Employing Eq. (\ref{52}) and taking into account (\ref{28}), we find
\begin{equation}\label{59}
\langle\hat{b}\hat{b}^{\dagger}\rangle={\gamma_{c}\over\kappa}\langle\hat{N}_{b}\rangle.
\end{equation}
Thus with the aid of (\ref{58}) along with (\ref{53}) and (\ref{59}), we arrive at
\begin{equation}\label{60}
(\Delta n)^{2}=\bigg({\gamma_{c}\over{\kappa}}\bigg)^{2}\langle\hat{N}_{a}\rangle\langle\hat{N}_{b}\rangle.
\end{equation}
We see from Eq. (\ref{40}) that for the two-level laser operating well above threshold
\begin{equation}\label{61}
\langle\hat{N}_{a}\rangle=N,
\end{equation}
so that on account of this and (\ref{23}), we arrive at
\begin{equation}\label{62}
\langle\hat{N}_{b}\rangle=0.
\end{equation}
Therefore, for $\gamma_{c}\ll r_{a}$, the variance of the photon number turns out to be
\begin{equation}\label{63}
(\Delta n)^{2}=0.
\end{equation}
This represents the normally-ordered variance of the photon number for coherent light.
On the other hand, for the same laser operating at threshold, we see that the variance of the photon number is
\begin{equation}\label{64}
(\Delta n)^{2}=\overline{n}^{2},
\end{equation}
which represents the normally-ordered variance of the photon number for chaotic light.

\section{Quadrature variance}
We next wish to calculate the quadrature variance for the plus and minus quadrature operators defined by
\begin{equation}\label{65}
\hat{b}_{+}=\hat{b}^{\dagger}+\hat{b}
\end{equation}
and
\begin{equation}\label{66}
\hat{b}_{-}=i(\hat{b}^{\dagger}-\hat{b}).
\end{equation}
It can be readily established that
\begin{equation}\label{67}
[\hat{b}_{-},\hat{b}_{+}]=2i{\gamma_{c}\over\kappa}(\hat{N}_{a}-\hat{N}_{b}).
\end{equation}
Thus on account of this result, we see that
\begin{equation}\label{68}
\Delta b_{+}\Delta b_{-}\geq{\gamma_{c}\over\kappa}\bigg|\langle\hat{N}_{a}\rangle-\langle\hat{N}_{b}\rangle\bigg|.
\end{equation}
Moreover, taking into account (\ref{53}) and (\ref{59}), one can easily verify that
\begin{equation}\label{69}
(\Delta b_{\pm})^{2}={\gamma_{c}\over\kappa}\bigg(\langle\hat{N}_{a}\rangle+\langle\hat{N}_{b}\rangle\bigg).
\end{equation}
Now in view of Eq. (\ref{61}) and (\ref{62}), we see that for the laser operating well above threshold
\begin{equation}\label{70}
(\Delta b_{+})^{2}=(\Delta b_{-})^{2}=\overline{n},
\end{equation}
with $\overline{n}$ is given by Eq. (\ref{55}).
In addition, for the laser operating at threshold, we have
\begin{equation}\label{71}
(\Delta b_{+})^{2}=(\Delta b_{-})^{2}=2\overline{n},
\end{equation}
in which $\overline{n}$ is given by (\ref{56}).
This represents the normally-ordered quadrature variance for chaotic light.

We define coherent light to be a light mode in which the uncertainties in the two quadratures are equal and satisfy the minimum uncertainty relation. On account of  Eq. (\ref{68}), we note that for the laser operating well above threshold,
$\Delta b_{+}\Delta b_{-}\geq\overline{n}$. Hence on the basis of this result and Eq. (\ref{70}), we assert that the light generated by the two-level laser operating well above threshold is coherent. We have seen that the quadrature variance for this case takes the relatively small value described by Eq. (\ref{70}). However, this value markedly differs from the corresponding value for an ideal coherent light.\footnote{For an ideal coherent light, the normally-ordered quadrature variance is identically zero.} This must be due to the fact that the coherent light generated by the two-level laser is represented by operators with vanishing mean. Perhaps it is also worth mentioning that whenever the laser coherent light is used as a driving or pump mode, we may replace the operators representing this light mode by the square root of the steady-state mean photon number. Such replacement transforms the laser coherent light into an ideal coherent light.

\section{Power spectrum}
It is also interesting to consider the power spectrum of the cavity light. The power spectrum of a single-mode light with central frequency $\omega_{0}$ is expressible as
\begin{equation}\label{72}
P(\omega)={1\over\pi}Re\int_{0}^{\infty}d\tau e^{i(\omega-\omega_{0})\tau}\langle\hat{b}^{\dagger}(t)
\hat{b}(t+\tau)\rangle_{ss}.
\end{equation}
Upon integrating both sides of Eq. (\ref{72}) over $\omega$, we readily get
\begin{equation}\label{73}
\int_{-\infty}^{\infty}P(\omega)d\omega=\overline{n},
\end{equation}
in which $\overline{n}$ is the steady-state mean photon number.
From this result, we observe that $P(\omega)d\omega$ is the steady-state mean photon number in the interval between $\omega$ and $\omega+d\omega$ [12].

We now proceed to calculate the two-time correlation function that appears in Eq. (\ref{72}) for the cavity light. To this end, we realize that the solution of Eq. (\ref{36}) can be written as
\begin{equation}\label{74}
\hat{b}(t+\tau)=\hat{b}(t)e^{-\kappa\tau/2}
+{g\over\sqrt{N}}e^{-\kappa\tau/2}
\int_{0}^{\tau}e^{\kappa\tau'/2}\hat{m}_{a}(t+\tau')d\tau',
\end{equation}
On the other hand, the solution of Eq. (\ref{42}) is expressible as
\begin{equation}\label{75}
\hat{m}_{a}(t+\tau)=\hat{m}_{a}(t)e^{-\eta\tau/2}
+e^{-\eta\tau/2}\int_{0}^{\tau}e^{\eta\tau''/2}\hat{F}(t+\tau'')d\tau'',
\end{equation}
so that on introducing this into Eq. (\ref{74}), we have
\begin{eqnarray}\label{76}
\hspace*{-12mm}\hat{b}(t+\tau)\hspace*{-3mm}&=&\hspace*{-3mm}\hat{b}(t)e^{-\kappa\tau/2}
+{g\over\sqrt{N}}e^{-\kappa\tau/2}\hat{m}_{a}(t)
\int_{0}^{\tau}e^{(\kappa-\eta)\tau'/2}d\tau'\nonumber\\&&
\hspace*{-12mm}+{g\over\sqrt{N}}e^{-\kappa\tau/2}\int_{0}^{\tau}d\tau'\int_{0}^{\tau'}d\tau'' e^{((\kappa-\eta)\tau'+\eta\tau'')/2}
\hat{F}(t+\tau'').
\end{eqnarray}
\begin{figure}[hbt]
\centering
\includegraphics[height=8cm,keepaspectratio]{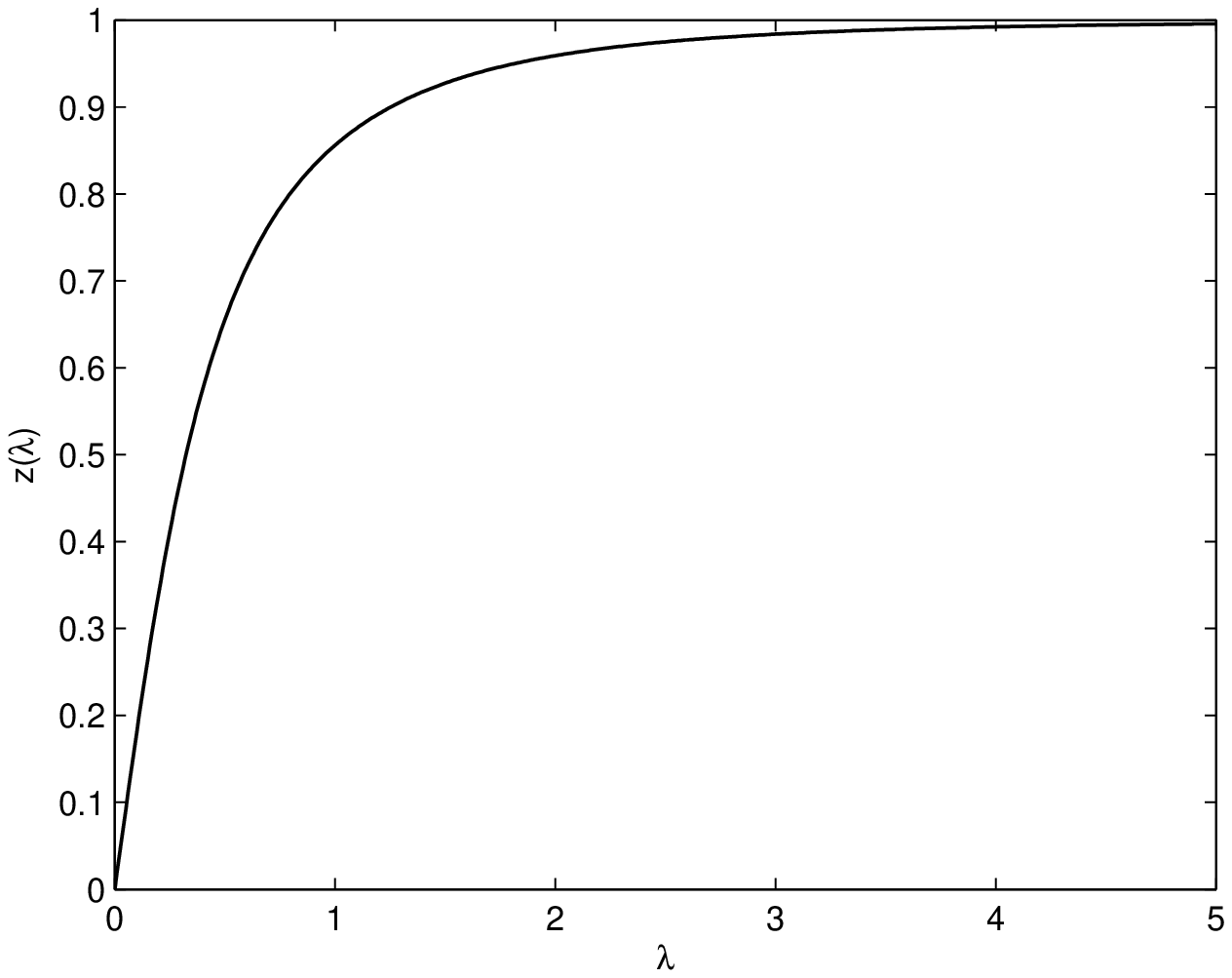}
\begin{center}
{\footnotesize{\bf Fig. 1}~~~~~A plot of Eq. (\ref{86}) for $\kappa=0.8$ and $\eta=5$.}
\end{center}
\end{figure}
\noindent
Thus on carrying out the first integration, we arrive at
\begin{eqnarray}\label{77}
\hat{b}(t+\tau)\hspace*{-3mm}&=&\hspace*{-3mm}\hat{b}(t)e^{-\kappa\tau/2}
+{2g\hat{m}_{a}(t)\over\sqrt{N}(\kappa-\eta)}\bigg[e^{-\eta\tau/2}-e^{-\kappa\tau/2}\bigg]\nonumber\\&&
+{g\over\sqrt{N}}e^{-\kappa\tau/2}\int_{0}^{\tau}d\tau'\int_{0}^{\tau'}d\tau'' e^{((\kappa-\eta)\tau'+\eta\tau'')/2}
\hat{F}(t+\tau'').
\end{eqnarray}
Now multiplying both sides on the left by $\hat{b}^{\dagger}(t)$ and taking the expectation value of the resulting equation, we have
\begin{eqnarray}\label{78}
\hspace*{-10mm}\langle\hat{b}^{\dagger}(t)\hat{b}(t+\tau)\rangle\hspace*{-3mm}&=&\hspace*{-3mm}\langle\hat{b}^{\dagger}(t)\hat{b}(t)\rangle e^{-\kappa\tau/2}+{2g\langle\hat{b}^{\dagger}(t)\hat{m}_{a}(t)\rangle\over\sqrt{N}(\kappa-\eta)}\bigg[e^{-\eta\tau/2}
-e^{-\kappa\tau/2}\bigg]\nonumber\\&&
\hspace*{-10mm}+{g\over\sqrt{N}}e^{-\kappa\tau/2}\int_{0}^{\tau}d\tau'\int_{0}^{\tau'}d\tau''e^{((\kappa-\eta)\tau'+\eta\tau'')/2}
\langle\hat{b}^{\dagger}(t)\hat{F}(t+\tau'')\rangle.
\end{eqnarray}

Furthermore, applying the adjoint of Eq. (\ref{52}) and taking into account (\ref{26}), we have
\begin{eqnarray}\label{79}
\hspace*{-10mm}\langle\hat{b}^{\dagger}(t)\hat{b}(t+\tau)\rangle\hspace*{-3mm}&=&\hspace*{-3mm}\langle\hat{b}^{\dagger}(t)\hat{b}(t)\rangle e^{-\kappa\tau/2}
+{\kappa\overline{n}\over(\kappa-\eta)}\bigg[e^{-\eta\tau/2}-e^{-\kappa\tau/2}\bigg]\nonumber\\&&
\hspace*{-10mm}+{\gamma_{c}\over 2N}e^{-\kappa\tau/2}\int_{0}^{\tau}d\tau'\int_{0}^{\tau'}d\tau''e^{((\kappa-\eta)\tau'+\eta\tau'')/2}
\times\langle\hat{m}_{a}^{\dagger}(t)\hat{F}(t+\tau'')\rangle,
\end{eqnarray}
so that in view of the fact that
\begin{equation}\label{80}
\langle\hat{m}_{a}^{\dagger}(t)\hat{F}(t+\tau'')\rangle=0,
\end{equation}
there follows
\begin{eqnarray}\label{81}
\langle\hat{b}^{\dagger}(t)\hat{b}(t+\tau)\rangle={\kappa\overline{n}\over{\kappa-\eta}}e^{-\eta\tau/2}
-{\eta\overline{n}\over{\kappa-\eta}}e^{-\kappa\tau/2}.
\end{eqnarray}
Finally, on combining (\ref{81}) with (\ref{72}) and carrying out the integration, we readily arrive at
\begin{eqnarray}\label{82}
P(\omega)\hspace{-3mm}&=\hspace{-3mm}&{\kappa\overline{n}\over{\kappa-\eta}}\bigg[{\eta/2\pi\over{(\omega-\omega_{0})^{2}
+[\eta/2]^{2}}}\bigg]
-{\eta\overline{n}\over{\kappa-\eta}}\bigg[{\kappa/2\pi\over{(\omega-\omega_{0})^{2}
+[\kappa/2]^{2}}}\bigg].
\end{eqnarray}

We realize that the mean photon number in the interval between $\omega'=-\lambda$ and $\omega'=\lambda$ is expressible as
\begin{equation}\label{83}
\overline{n}_{\pm\lambda}=\int_{-\lambda}^{\lambda}P(\omega')d\omega',
\end{equation}
in which $\omega'=\omega-\omega_{0}$. Therefore, upon substituting (\ref{82}) into Eq. (\ref{83}) and carrying out the integration, applying the relation
\begin{equation}\label{84}
\int_{-\lambda}^{\lambda}{dx\over x^{2}+a^{2}}={2\over a}tan^{-1}\bigg({\lambda\over a}\bigg),
\end{equation}
we arrive at
\begin{equation}\label{85}
\overline{n}_{\pm\lambda}=\overline{n}z(\lambda),
\end{equation}
where $z(\lambda)$ is given by
\begin{eqnarray}\label{86}
z(\lambda)={2\kappa/\pi\over{\kappa-\eta}}tan^{-1}\bigg({2\lambda\over\eta}\bigg)
-{2\eta/\pi\over{\kappa-\eta}}tan^{-1}\bigg({2\lambda\over\kappa}\bigg).
\end{eqnarray}
From the plot in Fig. 1, we easily find $z(0.5)=0.66$, $z(1)=0.86$, $z(2)=0.96$. Then combination of these results with Eq. (\ref{85}) yields $\overline{n}_{\pm 0.5}=0.66\overline{n}$, $\overline{n}_{\pm 1}=0.86\overline{n}$,
$\overline{n}_{\pm 2}=0.96\overline{n}$.

\section{conclusion}
We have carried out our analysis by putting the noise operators associated with the vacuum reservoir in normal order and applying the large-time approximation scheme. The procedure of normal ordering the noise operators renders the vacuum reservoir to be a noiseless physical entity. We uphold the viewpoint that the notion of a noiseless vacuum reservoir would turn out to be compatible with observation. Based on the definition of the stimulated emission decay constant, we infer that an atom in the upper level and inside a closed cavity emits a photon due to its interaction with the cavity light. We certainly identify this process to be stimulated emission.

We have found that for the two-level laser operating well above threshold, the uncertainties in the plus and minus quadratures are equal and satisfy the minimum uncertainty relation. In view of this, we have identified the light generated by the laser operating well above threshold to be coherent. The quadrature variance of the laser coherent light is markedly different from that of an ideal coherent light.\footnote{For an ideal coherent light, the normally-ordered quadrature variance is identically zero.} This must be due to the fact that the coherent light generated by the two-level laser is represented by operators with vanishing mean. Perhaps it is also worth mentioning that whenever the laser coherent light is used as a driving or pump mode, we may replace the operators representing this light mode by the square root of the steady-state mean photon number. Such replacement transforms the laser coherent light into an ideal coherent light. Moreover, we have seen that the light generated by the two-level laser when operating at threshold is chaotic and a large part of the total mean photon number is confined in a relatively small frequency interval.

\vspace*{5mm}
\noindent
{\bf References}
\vspace*{3mm}

\noindent
[1] D.F. Walls and J.G. Milburn, Quantum Optics (Springer-Verlag, Berlin, 1994).\newline
[2] M.O. Scully and M.S. Zubairy, Quantum Optics (Cambridge University Press, Cambridge,
\hspace*{5.5mm}1997).\newline
[3] R. Loudon, The Quantum Theory of Light (Oxford University Press, New York, 2000).\newline
[4] M.O. Scully and W.E. Lamb Jr., Phys. Rev. 208, 159 (1967).\newline
[5] Y.K. Wang and W.E. Lamb Jr., Phys. Rev. A 8, 866 (1973).\newline
[6] K.J. McNeil and D.F. Walls, J. Phys. A 8, 104 (1975).\newline
[7] F.T. Arecchi, M. Asdente, and A.M. Ricca, Phys. Rev. A 14, 383 (1976).\newline
[8] C. Benkert and O.M. Scully, Phys. Rev. A 41, 2756 (1990).\newline
[9] R. Pike and S. Sarkar, The Quantum Theory of Radiation (Oxford University Press, New York, \hspace*{5.2mm}1996).\newline
[10] L. Davidovich, Rev. Mod. Phys. 68, 127 (1997).\newline
[11] Fesseha Kassahun, Opt. Commun. 284, 1357 (2011).\newline
[12] Fesseha Kassahun, Fundamentals of Quantum Optics (Lulu Press Inc., North Carolina, 2010).

\end{document}